\documentclass[a4paper,11pt]{article}
\usepackage{jinstpub} % for details on the use of the package, please see the JINST-author-manual
\usepackage{lineno}
\usepackage{indentfirst}
\usepackage{graphicx}
\usepackage{subcaption}
\usepackage[export]{adjustbox}
\usepackage{wrapfig}
\usepackage{ulem}
\title{\boldmath Absolute value measurement of ion-scale turbulence by two-dimensional phase contrast imaging in Large Helical Device}

\author[a]{T. Kinoshita}
\author[b,c]{K. Tanaka}
\author[c]{H. Sakai}
\author[b]{R. Yanai}
\author[b,d]{M. Nunami}
\author[e]{C. A. Michael}
\affiliation[a]{Research Institute for Applied Mechanics, Kyushu University, Kasuga, 816-8580, Japan}
\affiliation[b]{National Institute for Fusion Science, National Institutes of Natural Sciences, Toki, 509-5292, Japan}
\affiliation[c]{Interdisciplinary Graduate School of Engineering Sciences, Kyushu University, Kasuga, 816-8580, Japan}
\affiliation[d]{Graduate School of Science, Nagoya University, Nagoya, 464-8603, Japan}
\affiliation[e]{Department of Physics and Astronomy, University of California - Los Angeles, Los Angeles, CA 90095-7099, United States of America}

\emailAdd{t.kinoshita@triam.kyushu-u.ac.jp}

\abstract{Absolute value measurements of turbulence amplitude in magnetically confined high-temperature plasmas can effectively explain turbulence-driven transport characteristics and their role in plasma confinements.
Two-dimensional phase contrast imaging (2D-PCI) is a technique to evaluate the space-time spectrum  of ion-scale electron density fluctuation.
However, absolute value measurement of turbulence amplitude has not been conducted owing to the nonlinearity of the detector.
In this study, the absolute measurement method proposed in the previous study is applied to turbulence measurement results in the large helical device.
As a result, the localized turbulence amplitude at $n_e=1.5\times 10^{19}$m$^{-3}$ is approximately $3.5\times 10^{15}$m$^{-3}$, which is 0.02\% of the electron density.
In addition, the evaluated poloidal wavenumber spectrum is almost consistent, within a certain error range, the spectrum being calculated using a nonlinear gyrokinetic simulation.
This result is the first to the best of our knowledge to quantitatively evaluate turbulence amplitudes measured by 2D-PCI and compare with simulations.
}

\keywords{phase contrast imaging, absolute measurement, turbulence, confinement, LHD}

\arxivnumber{2308.15713} % Only if you have one

\begin{document}
\maketitle
\flushbottom

\section{Introduction}
\label{sec:intro}
Energy and particle transport in magnetically confined plasmas usually exceeds the predictions of neoclassical transport, and the turbulence-driven anomalous transport plays an important role in the plasma confinement.
The relationships between the transport channels (particle flux $\Gamma_j$ and heat flux $Q_j$) and turbulence (density fluctuation $\tilde{n}_j$, temperature fluctuation $\tilde{T}_j$ and electric field fluctuation $\tilde{E}_\theta$ in the poloidal direction $\theta$) are described as follows\cite{wootton1990fluctuations}.
   \begin{equation}
   \label{Pflux_turb}
        \Gamma_j=\langle\tilde{E}_\theta \tilde{n}_j	\rangle / B_\phi
   \end{equation}
    \begin{equation}
   \label{hflux_turb}
        Q_j=\frac{3}{2} k_b n_j\langle\tilde{E}_\theta \tilde{T}_j	\rangle / B_\phi + \frac{3}{2} k_b T_j \langle\tilde{E}_\theta \tilde{n}_j	\rangle / B_\phi
   \end{equation}
Here, $n_j$, $T_j$, $B_\phi$, and $k_b$ correspond to the density, temperature of particle species $j$, magnetic field in the toroidal direction $\phi$, and Boltzmann constant, respectively.
The ensemble mean, denoted by $\langle\cdots\rangle$, indicates that each transport cannot be explained by only one of the fluctuations.
However, it is not practical to measure all kinds of fluctuations simultaneously, making it difficult to understand turbulence-driven transport in high-temperature plasmas.

Electron density fluctuation is a particularly important turbulence because it contributes to both transport channels. 
Moreover, it is relatively easy to measure, thus, various diagnostic techniques have been developed and installed to the high-temperature plasma devices\cite{brower1987spectrum, tokuzawa2021w, nakano2007reconstruction, fonck1990plasma, tokuzawa2012microwave, tanaka2008two}.
Phase contrast imaging (PCI) is an established technique to measure ion-scale electron density fluctuations\cite{tanaka2008two,weisen1988turbulent, coda1992phase, tanaka1993characteristics, tanaka2003phase, lin2009studies, edlund2018overview} and its space-time spectrum can be obtained employing a two-dimensional detector and magnetic shear method\cite{tanaka2008density, michael2008measurements}. 
The latter system is called two-dimensional phase contrast imaging (2D-PCI) and is in operation on the large helical device (LHD)\cite{tanaka2008density, michael2008measurements}.
Analysis methods of 2D-PCI to evaluate turbulence profile shape and absolute wave number have been established\cite{tanaka2008density, michael2008measurements}, while a method for the absolute measurement of amplitude has not been well established.
This is because the mercury cadmium tellurium (MCT) detector has nonlinear input-output (IO) characteristics, making absolute measurement of amplitude difficult.
Establishing an absolute measurement method for turbulence amplitude will enable comparisons of the turbulence characteristics via simulations as well as comparisons between high-temperature plasma devices, which will greatly contribute to further understanding turbulence-driven transport.

Recently, two evaluation methods have been proposed.
Z. Huang et al. proposed a method to acquire only the AC component, which is the scattered light due to turbulence, and estimate its absolute value by comparing it with known sound wave measurements\cite{huang2021wendelstein}.
On the other hand, we proposed a method to calibrate the detected AC and DC components (scattered and nonscattered light caused by turbulence) using the sensitivity curve of the detector and calculate absolute values\cite{kinoshita2020determination}.
In the previous study, we confirmed that the fluctuation amplitudes measured by PCI applying our proposed method are consistent with the interferometer measurements in a bench-top experiment\cite{kinoshita2020determination}.

In this study, the absolute value of the turbulence profile at LHD is evaluated applying our proposed method\cite{kinoshita2020determination} and comparing the results with those obtained via nonlinear gyrokinetic simulations.
The remainder of this paper is organized as follows: 
In Section 2, the evaluation of absolute value of line-integrated fluctuation is described. 
Evaluation of the absolute value of turbulence profile obtained from LHD is presented in Section 3. 
Section 4 describes the comparison of our results against those obtained via nonlinear gyrokinetic simulations. 
Finally, the results of this study are summarized in Sections 5.

\section{Evaluation of absolute value of line-integrated fluctuation}
\begin{figure}[b]
    \centering
    \includegraphics[width=13cm]{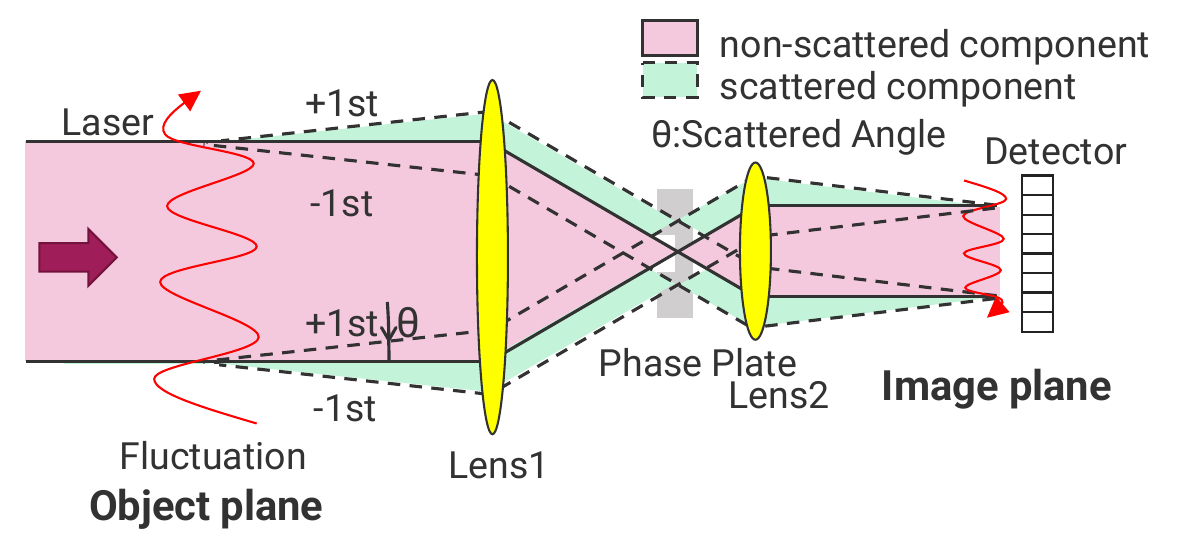}
    \caption{Principle of a fluctuation measurement conducted by PCI}
    \label{fig:fig_PCI_SVpci}
\end{figure}

In this section, our proposed method\cite{kinoshita2020determination} for evaluating the absolute value of the turbulence amplitude measured by PCI is described.
Figure \ref{fig:fig_PCI_SVpci} illustrates the principle of a fluctuation measurement by PCI. 
Electromagnetic wave propagating in turbulent plasmas is scattered by electron density fluctuations.
For instance, when a $\rm{CO_2}$ laser ($\lambda_i$=10.6$\rm{\mu}$m) passes through ion-scale electron density fluctuations, the density fluctuations act as optical gratings, producing positive and negative first order scattered lights (Raman-Nath diffraction).
Moreover, the phase of the scattered light is modulated due to the change in the refractive index caused by density fluctuations.
The amount of phase change $\tilde{\phi}$ corresponds to the amplitude of the fluctuation.
\begin{wrapfigure}[15]{r}{5.5cm}
  \begin{center}
    \raisebox{-4.5cm}[0pt][4.2cm]{\includegraphics[width=5cm]{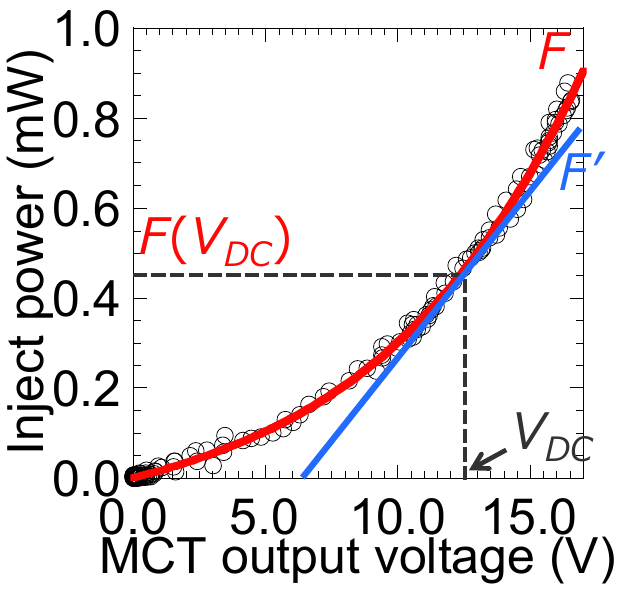}}
    \caption{I-O characteristics of 2D-MCT ch21}
    \label{fig:fig_PCI_BTSC}
  \end{center}
\end{wrapfigure}
The wave passing through the density fluctuation is expressed as
\begin{equation}
E=E_0e^{i\tilde{\phi}}\cong\left(1+i\tilde{\phi}\right)E_0,
    \label{eq:Ephase}
\end{equation}
where $E_0$ is the incident beam amplitude and $\tilde{\phi}$ is very small, thus, it can be replaced by the intensity change as expressed on the right-hand side.
Here, the first and second terms stated in parentheses on the right-hand side correspond to the nonscattered and scattered components, respectively.
However, in eq. (\ref{eq:Ephase}), the phase change involving fluctuation information cannot be detected because they are imaginary components.
Then, to convert an imaginary part into a real part, as expressed in eq. (\ref{eq:Ephase2}), a phase difference of $\pi$/2 rad. should be implemented between the nonscattered and scattered components, which is given by the phase plate.
\begin{equation}
E\cong\left(1+i\tilde{\phi}\right)E_0\rightarrow\left(1-\tilde{\phi}\right)E_0
    \label{eq:Ephase2}
\end{equation}
Finally, the nonscattered and scattered components are detected as intensity change as follows.
\begin{equation}
    I\cong{E_0}^2-2\tilde{\phi}{E_0}^2
\end{equation}
The first and second terms on the right-hand side correspond to the AC and DC components of the output voltage ($V_{AC}$ and $V_{DC}$) of a MCT detector.
The phase change $\tilde{\phi}$ can be obtained from $-V_{AC}/2V_{DC}$. 
However, a MCT detector has nonlinear I-O characteristics as shown in Figure \ref{fig:fig_PCI_BTSC}; therefore, it should be evaluated as follows.
\begin{equation}
        \tilde{\phi}=-\frac{F'(V_{DC})\times V_{AC}}{2F(V_{DC})}
        \label{eq:PCIACDC}
\end{equation}
Here, $F$ is a function of I-O characteristics as shown in Figure \ref{fig:fig_PCI_BTSC}. 
In a previous study, this method was applied to a bench-top experiment confirming that the fluctuation amplitudes evaluated by PCI were consistent with the interferometer measurements\cite{kinoshita2020determination}.
The conversion from phase change to line-integrated electron density fluctuation $\tilde{n}_eL$ is achieved as follows 
\begin{equation}
    \label{eq:PCInefluc}
        \tilde{n}_eL=\frac{4c^2\pi\epsilon_0m_e}{e^2}\frac{\tilde{\phi}}{\lambda_i},
\end{equation}
where $e$, $c$, $\epsilon_0$, and $m_e$ are the electric charge, speed of light, electric constant, and electron mass, respectively.

\section{Evaluation of absolute value of turbulence profile in LHD}
\subsection{Two-dimensional phase contrast imaging}
\begin{figure}[h]
  \begin{minipage}[h]{0.45\linewidth}
        \centering
        \includegraphics[width=4cm]{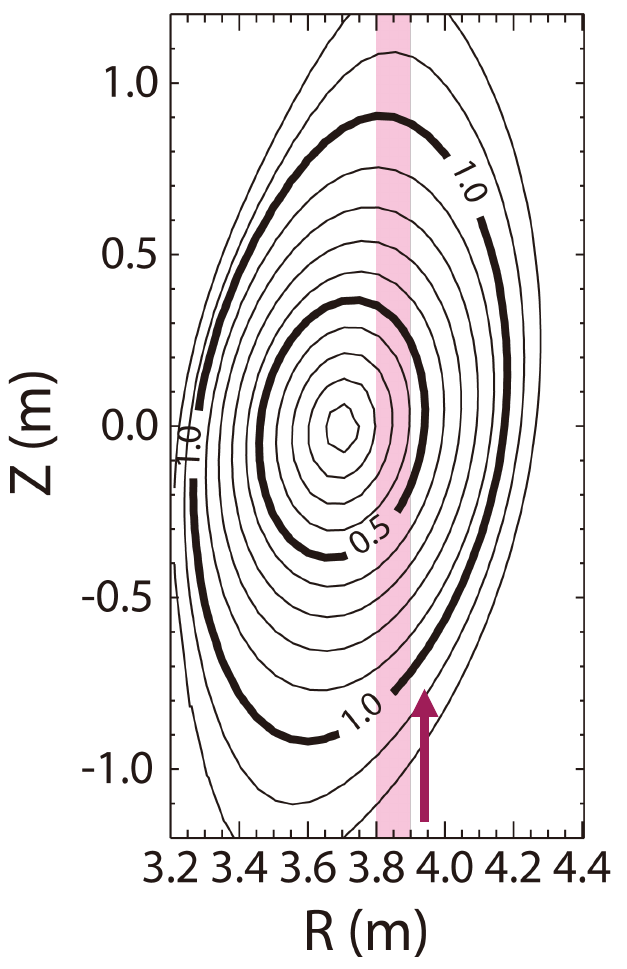}
        \caption{Line-of-sight of 2D-PCI in LHD}
        \label{fig:fig_PCI_LHD_CS}
  \end{minipage}
  \begin{minipage}[h]{0.5\linewidth}
       \centering
        \includegraphics[width=4.25cm]{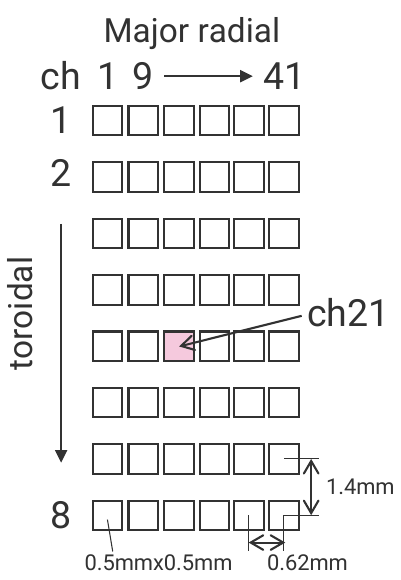}
        \caption{2D-MCT detector for 2D-PCI used in LHD}
        \label{fig:fig_PCI_LHD2DMCT}
  \end{minipage}
\end{figure}

Figure \ref{fig:fig_PCI_LHD_CS} shows the line-of-sight of 2D-PCI at the magnetic axis $R_{ax}$=3.6 m in LHD.
The electron density fluctuation measured by 2D-PCI is line-integrated along the line-of-sight.
To estimate the spatial profile, a two-dimensional MCT detector, as shown in Figure \ref{fig:fig_PCI_LHD2DMCT}, was employed and applied the magnetic shear technique\cite{tanaka2008two, michael2015two}.
The wavelength of the ion-scale turbulence is sufficiently long along the direction of the magnetic field lines and short along the direction perpendicular to the magnetic field lines.
Therefore, the coherence fluctuation components obtained from the two-dimensional correlation analysis between channels can be assumed to be a fluctuation propagating perpendicular to the magnetic field lines.
The magnetic shear method takes advantage of this characteristic to estimate the localized position based on the correspondence between the propagation direction and magnetic field line pitch angle.
The spatial resolution of turbulence profile is approximately 20\%. 

\subsection{Target plasma}
\begin{wrapfigure}[16]{r}[0pt]{6cm}
\vspace{-2\baselineskip}
  \begin{center}
    \raisebox{-7.2cm}[0pt][6.8cm]{\includegraphics[width=6cm]{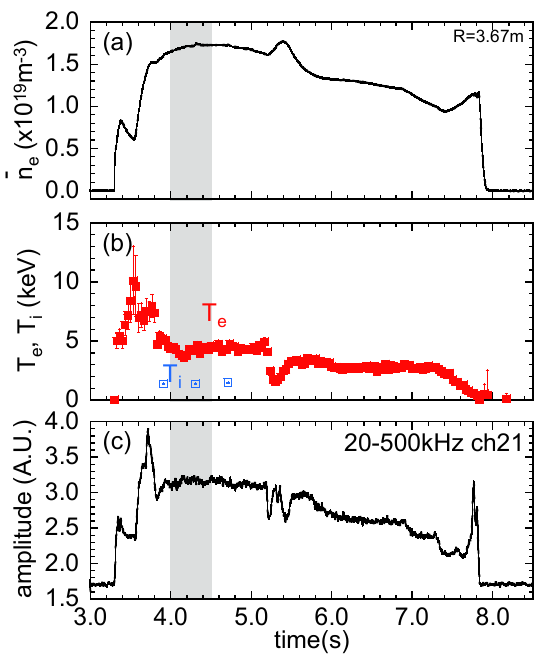}}
    \caption{Temporal evolution of (a) line-averaged electron density, (b) electron and ion temperatures, and (c) 2D-PCI AC signal amplitude (20-500 kHz, ch21), LHD experiment No.152264}
    \label{fig:fig_PCI_LHDwaveform}
  \end{center}
\end{wrapfigure}
Recently, a study comparing the transport in LHD and the Wendelstein 7-X (W7-X) was conducted\cite{warmer2021impact}.
At comparable electron density ($n_e=1.5\times 10^{19}$m$^{-3}$), the neoclassical ion-heat transport of W7-X is reported to be lower than that of LHD, and the anomalous ion-heat transport of LHD is lower than that of W7-X, resulting in comparable total heat transport\cite{warmer2021impact}.
Furthermore, the anomalous ion-heat flux is almost consistent with the ion-heat flux calculated by nonlinear gyrokinetic simulations (GKV\cite{watanabe2005velocity} for LHD and GENE\cite{jenko2000electron} for W7-X).
However, no comparison of the measured turbulence with nonlinear simulations or no direct comparison of the measured turbulence between LHD and W7-X has been reported.
This is because the evaluation method for absolute values of fluctuation amplitude was not established at that time.
In this study, the absolute value of the turbulence is evaluated for the LHD discharges used in this comparative study.
Figure \ref{fig:fig_PCI_LHDwaveform} shows a temporal evolution of (a) line-averaged electron density, (b) electron and ion temperatures, and (c) turbulence amplitude (a.u.).
We evaluate the absolute value of turbulence profile at the shaded region depicted in Figure \ref{fig:fig_PCI_LHDwaveform}.

\subsection{Process of evaluating absolute values of turbulence profile}
\begin{figure}[bp]
\centering
\includegraphics[width=10cm]{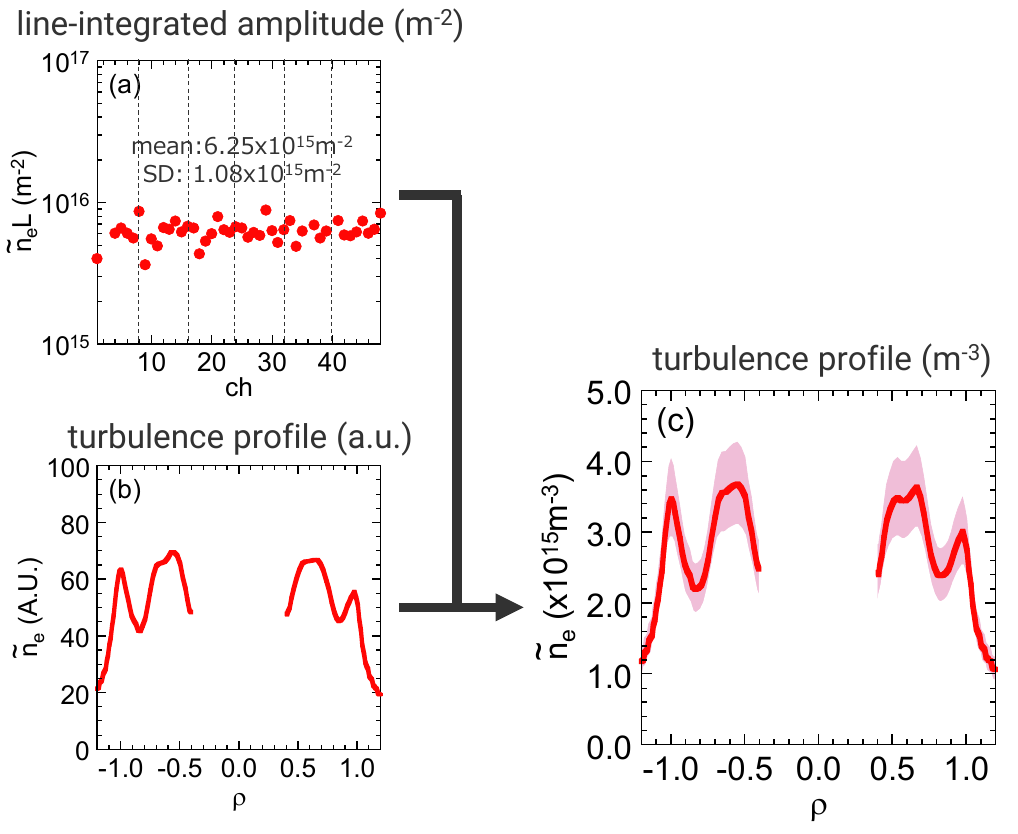}
\caption{Procedure to evaluate the absolute value of the electron density fluctuation distribution. The normalized radius $\rho$ is defined as the effective minor radius normalized by radius $a_{99}$, where the electron pressure inside $a_{99}$ is equal to 99\% of the total electron pressure \cite{suzuki2012development}.\label{fig:fig_PCI_LHDflucNL}}
\end{figure}
The absolute values of the turbulence profile is evaluated according to Figure \ref{fig:fig_PCI_LHDflucNL} and the following steps.
\begin{enumerate}
  \setlength{\parskip}{0cm} 
  \setlength{\itemsep}{0cm} 
   \item The absolute value of the line-integrated electron density fluctuation amplitude $\tilde{n}_eL$(m$^{-2}$) in each channel are evaluated as shown in Figure \ref{fig:fig_PCI_LHDflucNL}(a). Here, the effective value of the periodic fluctuation signal is assumed to be the amplitude of each channel. 
   \item The electron density fluctuation profile $\tilde{n}_e$(a.u.) is evaluated in arbitrary units using the magnetic shear technique as shown in Figure \ref{fig:fig_PCI_LHDflucNL}(b).
   \item The electron density fluctuation profile evaluated in arbitrary units is line-integrated along the line-of-sight. Then, the line-integrated electron density fluctuation amplitude $\tilde{n_e}L$(a.u.) is evaluated based on eq. (\ref{eq:PCInefluc}).   
   \item The absolute value conversion factor is obtained by dividing the average value among channels of $\tilde{n}_eL$(m$^{-2}$) by $\tilde{n}_eL$(a.u.).
   \item The conversion factor is multiplied by $\tilde{n}_e$(a.u.) to evaluate the absolute value $\tilde{n}_e$(m$^{-3}$) of the electron density fluctuation profile.
   \item The absolute calibration error is evaluated based on the standard deviation among channels of $\tilde{n}_eL$(m$^{-2}$).
\end{enumerate}
Here, as the beam diameter is sufficiently small compared to the plasma size, the turbulence amplitude should be constant among all channels (44/48 channels in operation) in the absence of local spatial structure. 
However, as shown in Figure \ref{fig:fig_PCI_LHDflucNL}(a), there is a variation of $\pm$15\%. 
In particular, it was observed around the dashed line depicted in Figure \ref{fig:fig_PCI_LHDflucNL}(a), which corresponds to the edge channels of the 2D-MCT detector array as shown in Figure \ref{fig:fig_PCI_LHD2DMCT}.
This is due to the weak laser intensity at the edge channels, which results in a weak scattered light intensity, and using a less accurate part of the sensitivity curve to convert for voltage-to-intensity conversion.

\begin{figure}[h]
\centering
\includegraphics[width=13cm]{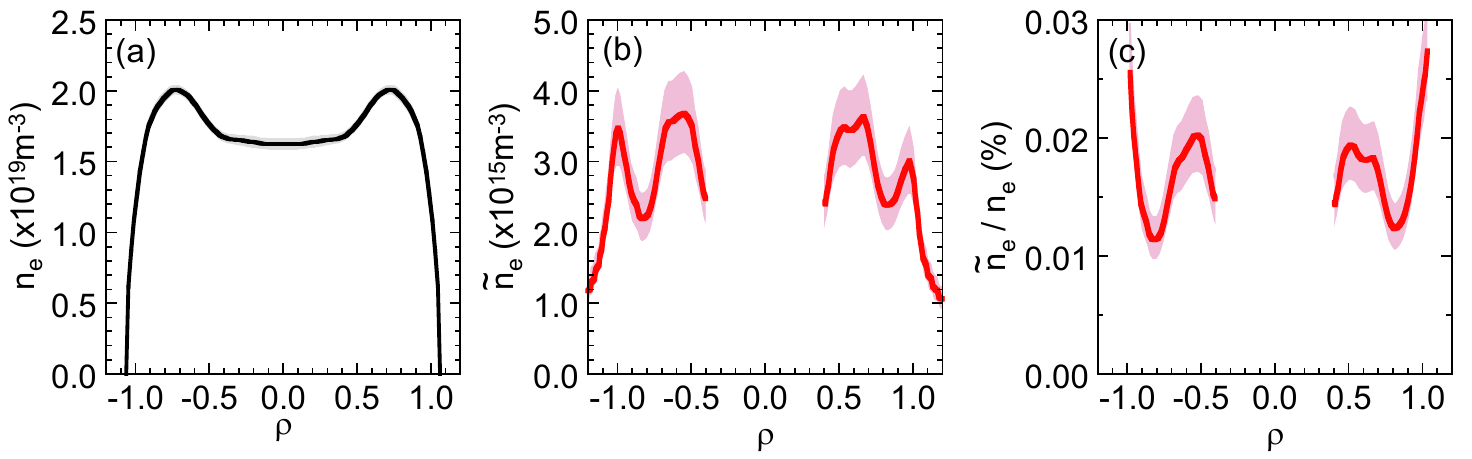}
\caption{Radial profiles of (a) electron density, (b) electron density fluctuation, and (c) normalized electron density fluctuation\label{fig:fig_PCI_LHDfluc_abs}}
\end{figure}

Figures \ref{fig:fig_PCI_LHDfluc_abs}(a)-(c) show the electron density profile, absolute value of the turbulence profile, and normalized turbulence profile, respectively.
The electron density profile was evaluated by applying the Abel transform to the far-infrared (FIR) laser interferometer results \cite{kawahata1999far, kawahata1997design, tanaka2008density}.
Figure \ref{fig:fig_PCI_LHDfluc_abs}(c) is (b) divided by (a) and depicts the magnitude of the fluctuation amplitude (fluctuation level) with respect to the electron density.
By applying the evaluation method proposed in this study to calibrate the output voltage based on the detector sensitivity and evaluating the absolute value of the ion-scale electron density fluctuation amplitude, we found that the localized fluctuation amplitude is approximately $3.5\times 10^{15}$m$^{-3}$, which is 0.02\% of the electron density.

\section{Comparison with nonlinear gyrokinetic simulation results }
A quantitative comparison of the turbulence levels evaluated as shown in Figure \ref{fig:fig_PCI_LHDfluc_abs}(c) and the calculated value by the nonlinear gyrokinetic simulation is presented in this section.
We carried out the local flux-tube nonlinear GKV\cite{watanabe2005velocity} code under following condition, the normalized ion temperature $R_0/L_{T_i}$=4.10 and density gradient $R_0/L_n$=-3.98, and the electron to ion temperature ratio $T_e/T_i$=1.99 at $t$=4.47s and $\rho$=0.5\cite{warmer2021impact}. 
Here, $L_X$ is the gradient scale length of a parameter X, $L_{X}$ = (-d$\ln$$X$/d$r$)$^{-1}$ and $R_0$ is the major radius, 3.63m. 

\begin{figure}[t]
  \begin{minipage}[h]{0.48\columnwidth}
        \centering
        \includegraphics[width=7cm]{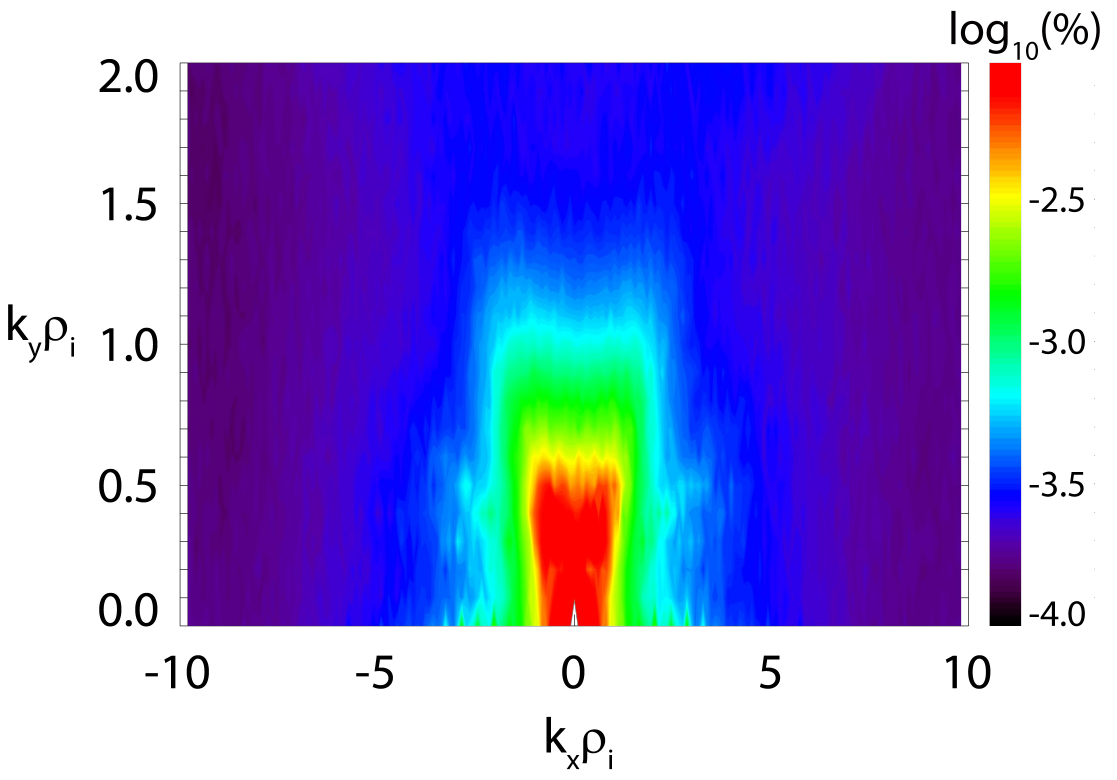}
        \caption{Two-dimensional wavenumber spectrum of the normalized electron density fluctuation.\label{fig:fig_PCI_LHDgkvkxky}}
  \end{minipage}
  \hspace{0.04\columnwidth}
  \begin{minipage}[h]{0.48\columnwidth}
        \centering
        \includegraphics[width=5cm]{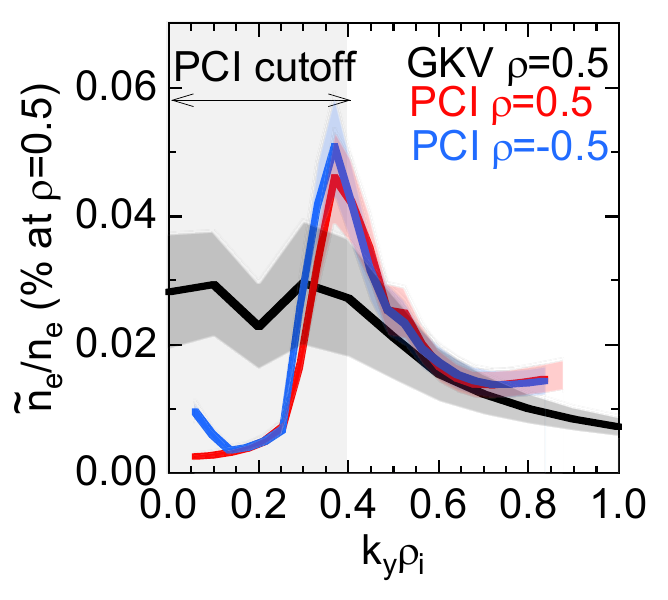}
        \caption{Comparison of poloidal wavenumber spectrum at $\rho=0.5$ between 2D-PCI and GKV values.\label{fig:fig_PCI_LHDgkvpciky}}
  \end{minipage}
\end{figure}

Figure \ref{fig:fig_PCI_LHDgkvkxky} is the two-dimensional wavenumber spectrum of the normalized electron density fluctuation calculated by GKV.
Here, $k_x\rho_i$ and $k_y\rho_i$ are toroidal and poloidal normalized wavenumbers, respectively.
Since PCI measures wavenumbers in the poloidal direction, a comparison is performed with the integral value along the $k_x\rho_i$ direction.
Figure \ref{fig:fig_PCI_LHDgkvpciky} shows a comparison of the wavenumber spectrum obtained by integrating Figure \ref{fig:fig_PCI_LHDgkvkxky} along the $k_x\rho_i$ direction over a wavenumber grid width and spectrum at $\rho$=$\pm$0.5 measured by 2D-PCI.
Here, solid lines and symbols denote the calculated values by GKV and experimental values measured by 2D-PCI, respectively.
Figure \ref{fig:fig_PCI_LHDgkvpciky} indicates that the measured values of 2D-PCI decrease significantly at $k_y\rho_i<$0.4, denoting that this region is the cutoff wavenumber for 2D-PCI in LHD\cite{kinoshita2020determination}.
However, for $0.5<k_y\rho_i<0.8$ (wavenumber measurable by 2D-PCI), the measured and calculated amplitude are almost consistent within the error bars, indicating that the evaluated absolute value in this study is reasonable.
This is the first quantitative comparison of turbulence spectrum predicted by nonlinear simulations and values measured by 2D-PCI.

\section{Summary}
In this study, we aimed to evaluate the absolute value of the ion-scale electron density fluctuation profile measured by 2D-PCI in LHD using the absolute value evaluation method proposed in a previous study.
The target plasma is ECRH plasma with $n_e=1.8\times 10^{19}$m$^{-3}$ used for comparison with W7-X results reported in a previous study.
Consequently, the electron density fluctuation amplitude in the ECRH plasma with $n_e=1.8\times 10^{19}$m$^{-3}$ is $3.5\times 10^{15}$m$^{-3}$, which is 0.02\% of the electron density.
In addition, the evaluated wavenumber spectrum at $\rho=\pm$0.5 is in good agreement with the spectrum calculated using nonlinear gyrokinetic simulations, indicating that the evaluated absolute value in this study is reasonable.
Unfortunately, 2D-MCT is hard to come by.
Therefore, a method to obtain a spatial profile by rotating the radial localization mask is implemented \cite{huang2021wendelstein}.
The evaluation of absolute values of the electron density fluctuation profile and wavenumber spectrum allows comparison with the simulation results presented in this study as well as direct comparison of turbulence between high-temperature plasma devices such as LHD and W7-X.

\acknowledgments
%LHD data can be accessed from the LHD data repository \url{https://www-lhd.nifs.ac.jp/pub/Repository_en.html}.
The data supporting the results of this study are available from the LHD experiment data repository at \url{https://doi.org/10.57451/lhd.analyzed-data}.
The authors thank the LHD experiment group for their excellent work on the operation of LHD. 
This study was supported by NIFS (17ULHH013, 18ULHH013, 19ULHH013, 20ULHH013, 21ULHH013, and 22ULHH013) and JSPS (21J12314, 16H04620, and 21H04458) grants.

% Bibliography

\bibliographystyle{JHEP}
\bibliography{biblio.bib}

\end{document}